\documentstyle[amssymb,aps,twocolumn]{revtex}
\begin{document}
\author{Jietai Jing, Jing Zhang, Ying Yan, Fagang Zhao, Changde Xie, Kunchi Peng}
\title{Experimental demonstration of tripartite entanglement and controlled dense
coding for continuous variables}
\address{The State Key Laboratory of Quantum Optics and Quantum Optics\\
Devices, Institute of Opto-Electronics, Shanxi University, Taiyuan, 030006,\\
P.R.China}
\maketitle

\begin{abstract}
A tripartite entangled state of bright optical field is experimentally
produced using an Einstein-Podolsky-Rosen entangled state for continuous
variables and linear optics. The controlled dense coding among a sender, a
receiver and a controller is demonstrated by exploiting the tripartite
entanglement. The obtained three-mode position correlation and relative
momentum correlation between the sender and the receiver and thus the
improvements of the measured signal to noise ratios of amplitude and phase
signals with respect to the shot noise limit are 3.28dB and 3.18dB
respectively. If the mean photon number $\overline{n}$ equals 11 the channel
capacity can be controllably inverted between 2.91 and 3.14. When $\overline{%
n}$ is larger than 1.0 and 10.52 the channel capacities of the controlled
dense coding exceed the ideal single channel capacities of coherent and
squeezed state light communication.

PACS numbers(s): 42.50.Dv, 03.67.Hk, 03.65.Ta
\end{abstract}

Quantum entanglement shared by more than two parties is the essential base
for developing quantum communication network and quantum computation. The
three-particle entangled states for discrete variables, called also
Greenberger-Horne-Zeilinger (GHZ) states, have been proposed\cite{1} and
then experimentally realized with optical system consisting of nonlinear ($%
\chi ^2$) crystal, pulse laser and linear optical elements\cite{2} and with
nuclear magnetic resonance\cite{3}. The controlled dense coding for discrete
variables using a three-particle entangled state has been proposed\cite{4}.
Recently, under the motivation of the successful experiments on
continuous-variable quantum teleportation\cite{5} and quantum dense coding%
\cite{6}, the schemes demonstrating quantum teleportation network\cite{7}
and controlled dense coding\cite{8} for quantum variables with a continuous
spectrum using multipartite entanglement have been theoretically proposed.
So far to the best of our knowledges, the experimental report on the
generation of multipartite entangled states for continuous variables and its
application has not been presented.

In this paper we report the first experimental demonstration of quantum
entanglement among more than two quantum systems with continuous spectra.
The tripartite entangled state is produced by distributing a two-mode
squeezed state light to three parties using linear optics. The obtained
tripartite entangled optical beams are distributed to a sender (Alice), a
receiver (Bob) and a controller (Claire) respectively. The information
transmission capacity of the quantum channel between Alice and Bob is
controlled by Claire. The channel capacity accomplished under Claire's help
is always larger than that without his help. For the large mean photon
number($\overline{n}>10.52$), the channel capacity of the controlled dense
coding communication exceeds that of ideal squeezed state communication.

Fig.1 is the schematic of the experimental setup for tripartite entanglement
generation and controlled dense coding. A semimonolithic nondegenerate
optical parameter amplifier (NOPA) involving an $\alpha $--cut type-II KTP
crystal and pumped by an intracavity frequency-doubled and
frequency-stabilized Nd:YAP/KTP laser serves as the initial bipartite
entanglement source. The configuration and operation principle of this
source have been detailedly described in our previous publications\cite{6,9}%
. The output optical modes with horizontal and vertical polarizations from
the NOPA operating at deamplification, $\widehat{b}_1$ and $\widehat{b}_2$ ,
are a pair of bright Einstein-Podolsky-Rosen(EPR) entangled beams with
anticorrelated amplitude quadratures and correlated phase quadratures\cite{6}%
. The polarizations of $\widehat{b}_1$ and $\widehat{b}_2$ are rotated by a
half-wave plate ($\lambda $/2) the optical axis of which is in $\theta =45^0-%
\frac 12arcsin(\frac{\sqrt{2}-1}{\sqrt{6}})$ relative to the horizontal
direction, then the beams pass through a polarizing-beam-splitter (PBS) with
horizontal and vertical polarizations. The output beam $\widehat{b}%
_2^{\prime }$ is splitted again by a 50/50 beam-splitter consisting of a
half-wave plate ($\lambda /2$) and a PBS to modes $\widehat{c}_2$ and $%
\widehat{c}_3$. In Ref.[8] we have proved theoretically that the modes $%
\widehat{c}_1,$ $\widehat{c}_2$ and $\widehat{c}_3$ are in a tripartite
entangled state which is a ''three-mode position eigenstate'' with the
quantum correlations of total position quadratures($\widehat{X}_{\widehat{c}%
_1},$ $\widehat{X}_{\widehat{c}_2}$and $\widehat{X}_{\widehat{c}_3}$)and
relative momentum quadratures ($\widehat{Y}_{\widehat{c}_1},$ $\widehat{Y}_{%
\widehat{c}_2}$and $\widehat{Y}_{\widehat{c}_3}$)(see Fig.4 of Ref.8 for the
case of $r_2=0$). The outgoing tripartite entangled state for continuous
variables, a bright ''GHZ-like'' state is utilized to implement the
controlled dense coding.

The entangled beams $\widehat{c}_1,$ $\widehat{c}_2$ and $\widehat{c}_3$ are
sent to Alice ,Bob and Clarie, respectively. Alice modulates two sets of
classical signals, which she wants to send to Bob, on the amplitude and
phase quadratures of her mode $\widehat{c}_1$by amplitude and phase
modulators AM and PM .For example, a specific encoding scheme of binary
pulse code modulation, in which the data are independently encoded as two
trains of 1 and 0 pulse signals at some radio frequency (rf)on the amplitude
and phase quadratures, can be applied. The modulations on mode $\widehat{c}%
_1 $lead to a displacement of $a_s:$%
\begin{equation}
\widehat{c}_1^{\prime }=\hspace*{0in}\widehat{c}_1+a_s,
\end{equation}
where $a_s=X_s+iY_s$ is the sent signal via the quantum channel. The
outgoing mode $\widehat{c}_1^{\prime }$is sent to Bob who imposes a phase
difference of $\pi $/2 between $\widehat{c}_1^{\prime }$and himself mode $%
\widehat{c}_2$with a phase-shifter(PS), and then combines the two modes on a
50/50 beam-splitter consisting of two PBS and a $\lambda /2$ wave-plate. The
two bright output beams from BS2 are directly detected by photodiodes D$_1$
and D$_2$. The each photocurrent of D$_1$ and D$_2$ is divided into two
parts with power splitters RF1 and RF2. Through analogous calculation with
that used in Ref.[8] but taking into account the imperfect detection
efficiency of the detectors($\eta <1\ $for D$_1$,D$_2$ and D$_3$) and the
nonzero losses of optical systems($\xi _1\neq 0$ for $\widehat{c}_1$and $%
\widehat{c}_2,\xi _2\neq 0$ for $\widehat{c}_3$) the noise power spectra of
the sum and difference photocurrents are expressed by\cite{10}: 
\begin{eqnarray}
\left\langle \delta ^2\widehat{i}_{+}\right\rangle &=&\eta ^2\xi _1^2\frac{%
e^{2r}+8e^{-2r}-9}{12}+1+\frac 12V_{X_s} \\
\left\langle \delta ^2\widehat{i}_{-}\right\rangle &=&3\eta ^2\xi _1^2\frac{%
e^{-2r}-1}4+1+\frac 12V_{Y_s}  \nonumber
\end{eqnarray}
where $r$ is the squeezing parameter of the EPR beams ($0\leqslant r<\infty $%
), $V_{X_s}$ and $V_{Y_s}$ are the fluctuation variances of the modulated
signals ($X_s,Y_s$). We can see from Eq.(2), for $r>0$,the quantum
fluctuation $\left\langle \delta ^2\widehat{i}_{+}\right\rangle $ is always
larger than $\left\langle \delta ^2\widehat{i}_{-}\right\rangle $, and the
larger the $r$ is, the bigger the deviation between $\left\langle \delta ^2%
\widehat{i}_{+}\right\rangle $ and $\left\langle \delta ^2\widehat{i}%
_{-}\right\rangle $ is. If $r\rightarrow \infty $, Bob only can gain the
phase signal with high accuracy beyond the shot noise limit (SNL), however,
he can not gain the amplitude signal that is submerged in large noise. For
extracting the amplitude signal Bob have to have the Claire's result of the
amplitude-quadrature detection. Claire detects the amplitude quadrature of
mode $\widehat{c}_3$with photodiode D$_3$ and sends the measured
photocurrent to Bob. Bob displaces the Claire's result on the sum
photocurrent : 
\begin{eqnarray}
\widehat{i}_{+}^{\prime } &=&\widehat{i}_{+}+g\widehat{i}_3  \nonumber \\
&=&\frac{\eta \xi _1}{\sqrt{2}}(\widehat{X}_{\widehat{c}_1^{\prime }}+%
\widehat{X}_{\widehat{c}_2})+g\eta \frac{\xi _2^2}{\xi _1}\widehat{X}_{%
\widehat{c}_3}+\frac{\eta \sqrt{1-\xi _1^2}}{\sqrt{2}}(\widehat{X}_{v_{%
\widehat{c}_1}}  \nonumber \\
&&+\widehat{X}_{v_{\widehat{c}_2}})+\frac{\sqrt{1-\eta ^2}}2(\widehat{X}%
_{v_{D_1}}+\widehat{Y}_{v_{D_1}}+\widehat{X}_{v_{D_2}}-\widehat{Y}_{v_{D_2}})
\\
&&+\frac{g\eta \xi _2^2\sqrt{1-\xi _2^2}}{\xi _1}\widehat{X}_{v_{\widehat{c}%
_3}}+\frac{g\xi _2\sqrt{1-\eta ^2}}{\xi _1}\widehat{X}_{v_{D_3}}+\frac 1{%
\sqrt{2}}X_s,  \nonumber
\end{eqnarray}
where $g$ describes gain at Bob for the transformation from Claire's
photocurrent to his sum photocurrent. The optimal gain for attaining the
minimum variances of the sum photocurrent is $g_{opt}=\frac{%
(e^{4r}+3e^{2r}-4)\eta ^2\xi _1^2}{\sqrt{2}(e^{4r}\eta ^2\xi
_2^2-3e^{2r}\eta ^2\xi _2^2+6e^{2r}+2\eta ^2\xi _2^2)}$. It is easy to be
seen, for larger squeezing the optimum gain of the sum photocurrent is $g=%
\frac 1{\sqrt{2}}$. For simplification and without losing generality, we
take $g=\frac 1{\sqrt{2}}$ in the following calculation and experiment, so
the power fluctuation spectrum of sum photocurrent of three modes equals to
: 
\begin{eqnarray}
\left\langle \delta ^2\widehat{i}_{+}^{\prime }\right\rangle &=&\frac 1{12}%
\{e^{2r}\eta ^2(\frac{\xi _2^2-\xi _1^2}{\xi _1})^2+2e^{-2r}\eta ^2(\frac{%
\xi _2^2+2\xi _1^2}{\xi _1})^2  \nonumber \\
&&-3(\frac{\xi _2^4}{\xi _1^2}\eta ^2-4+3\eta ^2\xi _1^2+2\xi _2^2\eta ^2-2%
\frac{\xi _2^2}{\xi _1^2})\} \\
&&+\frac 12V_{X_s}.  \nonumber
\end{eqnarray}
Eq.(4) shows that Bob can also gain the amplitude signal with a sensitivity
beyond SNL under the help of Claire.

Fig.2(a) is the measured noise power spectra of the amplitude sums $%
\left\langle \delta ^2\widehat{i}_{+}^{\prime }\right\rangle $(trace 3) and $%
\left\langle \delta ^2\widehat{i}_{+}\right\rangle $(trace 2). In trace 2
the modulated amplitude signal at 2MHz is submerged in the noise floor of $%
\left\langle \delta ^2\widehat{i}_{+}\right\rangle $ and in trace 3 the
modulated signal emerges from the reduced noise floor of $\left\langle
\delta ^2\widehat{i}_{+}^{\prime }\right\rangle $ due to the help of Claire.
The measured noise floor of trace 3 is $1.57$dB lower than that of trace 2.
After the correction to the electronics noise floor(trace4) which is $\sim
7.83$dB below the SNL(trace1), the noise reductions of $(\widehat{X}_{%
\widehat{c}_1}+\hspace{0in}\widehat{X}_{\widehat{c}_2}+\widehat{X}_{\widehat{%
c}_3})$ and $(\widehat{X}_{\widehat{c}_1}+\widehat{X}_{\widehat{c}_2})$
relative to SNL should actually be $3.28$dB and 1.19dB(The difference of
1.57dB should be corrected to 2.09dB). The results definitely confirm the
existence of the quantum correlation among three amplitude quadratures of
modes $\widehat{c}_1,$ $\widehat{c}_2$ and $\widehat{c}_3$. The signal
modulated on the phase quadrature at 2MHz is detected by Bob directly\cite
{10}. The noise power spectrum of phase quadratures between the pair of $%
\widehat{c}_1$ and $\widehat{c}_2$ modes is shown in Fig.2(b). Trace 2 is
the measured noise power spectrum of ($\widehat{Y}_{\widehat{c}_1}-$ $%
\widehat{Y}_{\widehat{c}_2}$) which is $2.66$dB below the SNL(trace 1).
Accounting for the electronics noise (trace 3), it should be $3.18$dB below
the SNL actually. Substituting the measured noise power of $\left\langle
\delta ^2\widehat{i}_{+}\right\rangle ,\left\langle \delta ^2\widehat{i}%
_{+}^{\prime }\right\rangle $ and $\left\langle \delta ^2\widehat{i}%
_{-}\right\rangle $ from Fig.2 into the Eqs.(2) and (4), we calculate the
squeezing parameter $r_{\exp }=0.674$(5.85dB squeezing after the
correction)(The parameters $\left\langle \delta ^2\widehat{i}%
_{+}\right\rangle =0.76$, $\left\langle \delta ^2\widehat{i}_{+}^{\prime
}\right\rangle =0.47$, $\left\langle \delta ^2\widehat{i}_{-}\right\rangle
=0.48$, $\xi _1^2=98.7\%$, $\xi _2^2=93.7\%$, $\eta ^2=95.0\%$ are taken in
the calculation according to the experimental values).

Fig.3 is the functions of the normalized fluctuation variances of $%
\left\langle \delta ^2\widehat{i}_{-}\right\rangle $, $\left\langle \delta ^2%
\widehat{i}_{+}\right\rangle $ and $\left\langle \delta ^2\widehat{i}%
_{+}^{\prime }\right\rangle $ versus the squeezing parameter $r$, where $\xi
_1,\xi _2,\eta $ and $g=\frac 1{\sqrt{2}}$ are the values for the
experimental system. $\left\langle \delta ^2\widehat{i}_{+}^{\prime
}\right\rangle _{opt}$ is the fluctuation variance of the amplitude sum of
three modes when the optimal gain $g_{opt}$ is applied. We can see, the
difference between $\left\langle \delta ^2\widehat{i}_{+}^{\prime
}\right\rangle _{opt}$ and $\left\langle \delta ^2\widehat{i}_{+}^{\prime
}\right\rangle $ is quite small (0.035) for the experimental squeezing $%
r_{\exp }=0.674$ and the difference tends to zero for larger $r.$ $%
\left\langle \delta ^2\widehat{i}_{+}^{\prime }\right\rangle $ is smaller
than $\left\langle \delta ^2\widehat{i}_{+}\right\rangle $ and increasing $r$
, $\left\langle \delta ^2\widehat{i}_{+}\right\rangle $ increases, however $%
\left\langle \delta ^2\widehat{i}_{+}^{\prime }\right\rangle $ decreases.
The results clearly exhibit the tripartite entanglement among $\widehat{c}%
_1, $ $\widehat{c}_2$ and $\widehat{c}_3$ modes. When $r_{\exp }=0.674,$ $%
\left\langle \delta ^2\widehat{i}_{+}^{\prime }\right\rangle $ is 0.29 lower
than $\left\langle \delta ^2\widehat{i}_{+}\right\rangle .$ As shown in
Eq.(2) and Fig.(3), generally, the noise floor of amplitude signal is not
equal to that of phase signal, which is because the discussed state is not a
maximal GHZ state and decoding amplitude signal must have the aid of
Claire's classical information and phase signal only needs the joint
measurement of $\widehat{c}_1$ and $\widehat{c}_2$ modes.

Following the theoretical calculations on the quantum channel capacity for
dense coding in Ref.[8][11][12] we calculate the channel capacity of the
presented experimental system. The channel capacities with and without
Claire's help can be deduced from Eqs.(2) and (4):

\begin{eqnarray}
C_{n-c}^{dense} &=&\frac 12\ln [(1+\frac{\sigma ^2}{\left\langle \delta ^2%
\widehat{i}_{-}\right\rangle })(1+\frac{\sigma ^2}{\left\langle \delta ^2%
\widehat{i}_{+}\right\rangle })] \\
C_c^{dense} &=&\frac 12\ln [(1+\frac{\sigma ^2}{\left\langle \delta ^2%
\widehat{i}_{-}\right\rangle })(1+\frac{\sigma ^2}{\left\langle \delta ^2%
\widehat{i}_{+}^{\prime }\right\rangle })]  \nonumber
\end{eqnarray}
where, $\sigma ^2$ is the average value of the signal photon number and the
mean photon number per mode $\stackrel{-}{n}=\sigma ^2+\sinh ^2r$\cite{8,12}%
. The dependences of the channel capacities for ideal single mode coherent
state ($C^{ch}=\ln (1+\stackrel{-}{n})$) and squeezing state ($C^{sq}=\ln (1+%
\stackrel{-}{2n})$)\cite{8,11,12} on the mean photon number $\stackrel{-}{n}$
are given in Fig.4 to compare with that of controlled dense coding with ($%
C_c^{dense})$ and without ($C_{n-c}^{dense}$) Claire's help according to
Eq.5 and taking the experimental parameters. For the given squeezing ($%
r_{\exp }=0.674)$, when the mean photon number $\stackrel{-}{n}$ is larger
than 1.00(1.31) and 10.52 the channel capacities of the controlled dense
coding with(without) the help of Claire exceed that of coherent state and
squeezed state communication. By increasing the average signal photon number 
$\sigma ^2,$ the channel capacity of quantum dense coding can be improved%
\cite{11}. The channel capacity with the help of Claire ($C_c^{dense})$ is
always larger than that without his help ($C_{n-c}^{dense}$) which is just
the result of three-partite entanglement. For example, when $\overline{n}%
=11, $ the channel capacity of the presented system can be controllably
inverted between 2.91 and 3.14.

In conclusion, we experimentally obtained bright tripartite entangled state
light and accomplished the quantum controlled dense coding for the
continuous variables. We deduced the formulae designating the tripartite
entanglement among amplitude and phase quadratures of three modes in the
case of accounting for the influences of imperfect detection efficiency and
the losses of optical system and using gain $g=\frac 1{\sqrt{2}}$. The
experiment shows that using the limited squeezing the channel capacity of
the controlled dense coding can exceed that of coherent state and squeezed
state communication when the signal photon number is larger than a certain
value. The mature technique of optical parametric amplification, the simple
linear optical system and the direct measurement for Bell state are used in
the presented scheme, thus the experimental implementation is significantly
simplified. The presented schemes generating tripartite entanglement and
achieving controlled dense coding are valuable for developing future
information network of quantum systems with continuous spectra.

This research was supported by the National Fundamental Research Natural
Science Foundation of China (Grant No.2001CB309304), the National Natural
Science Foundation of China (Grant. No. 60238010, 60178012) and the Shanxi
Province Young Science Foundation (Grant No.20021014).

Caption

Fig.1 Experimental setup for tripartite entanglement generation and
controlled dense coding.

Fig.2(a) The noise power spectra of amplitude sums $\left\langle \delta ^2%
\widehat{i}_{+}^{\prime }\right\rangle $(trace 3) and $\left\langle \delta ^2%
\widehat{i}_{+}\right\rangle $(trace 2), trace 1---shot noise limit(SNL),
trace 4---Electronics noise level(ENL), measured frequency range:
1.5MHz-2.5MHz, resolution bandwidth 30KHz, video bandwidth 0.1KHz.

Fig.2(b) The noise power spectra of phase difference $\left\langle \delta ^2%
\widehat{i}_{-}\right\rangle $(trace 2), trace 1---SNL, trace 3---ENL,
measured frequency range: 1.5MHz-2.5MHz, resolution bandwidth 30KHz, video
bandwidth 0.1KHz.

Fig.3 The variances of amplitude sums $\left\langle \delta ^2\widehat{i}%
_{+}\right\rangle ,\left\langle \delta ^2\widehat{i}_{+}^{\prime
}\right\rangle $ $\left\langle \delta ^2\widehat{i}_{+}^{\prime
}\right\rangle _{opt}$ and phase difference $\left\langle \delta ^2\widehat{i%
}_{-}\right\rangle $ versus the squeezing parameter $r$ with beam
propagation efficiency $\xi _1^2=98.7\%$, $\xi _2^2=93.7\%$, and the quantum
efficiency of detector $\eta ^2=95.0\%$.

Fig.4 Channel Capacities for the controlled dense coding with ($C_c^{dense}$%
) and without ($C_{n-c}^{dense}$) Claire's help, single-mode coherent state
with heterodyne detection , and squeezed state ($C^{sq}$) communication. The
parameters are same with Fig.3.

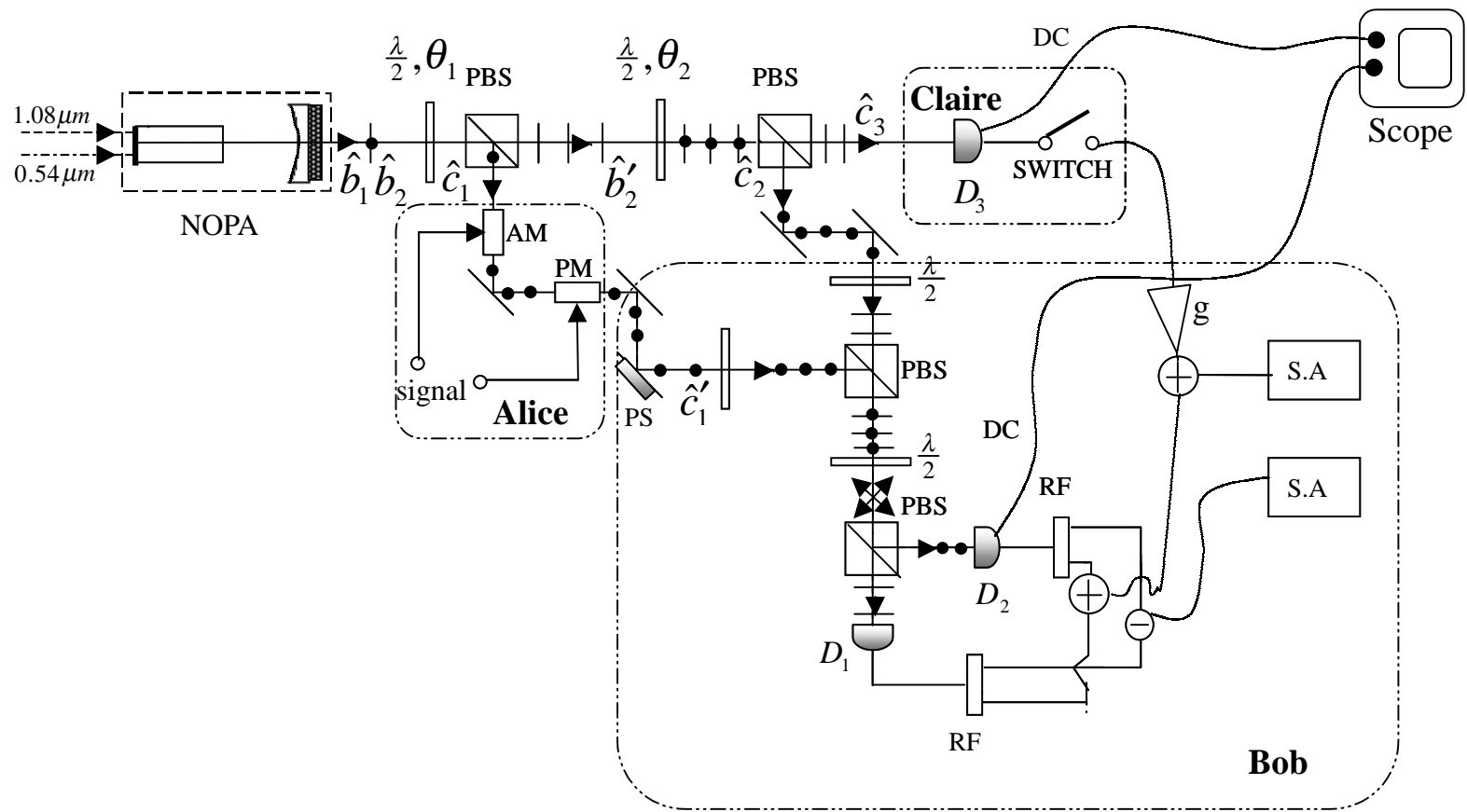
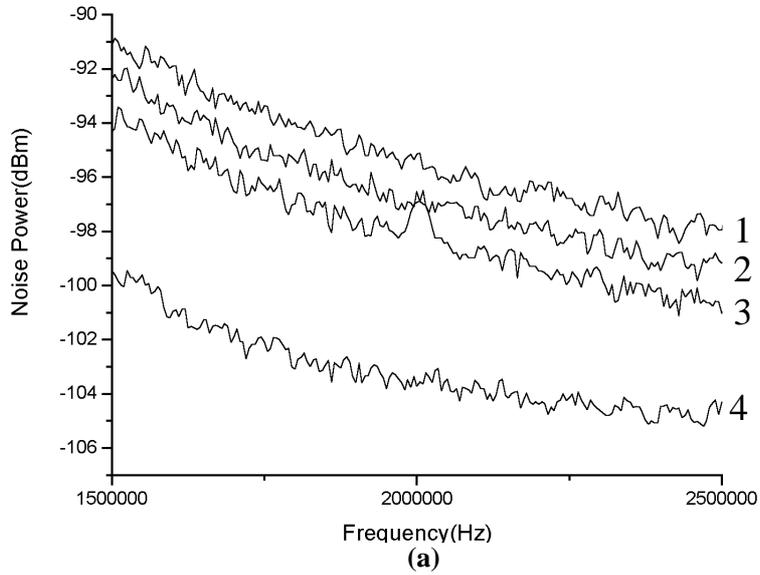
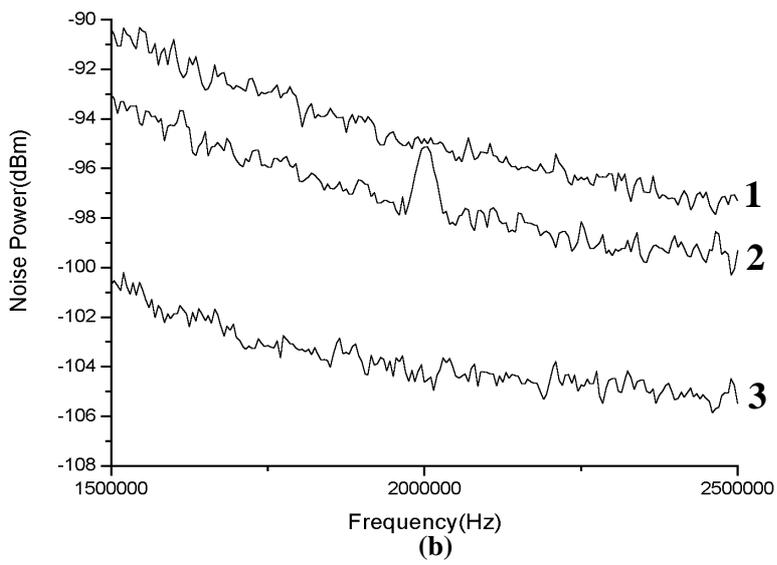
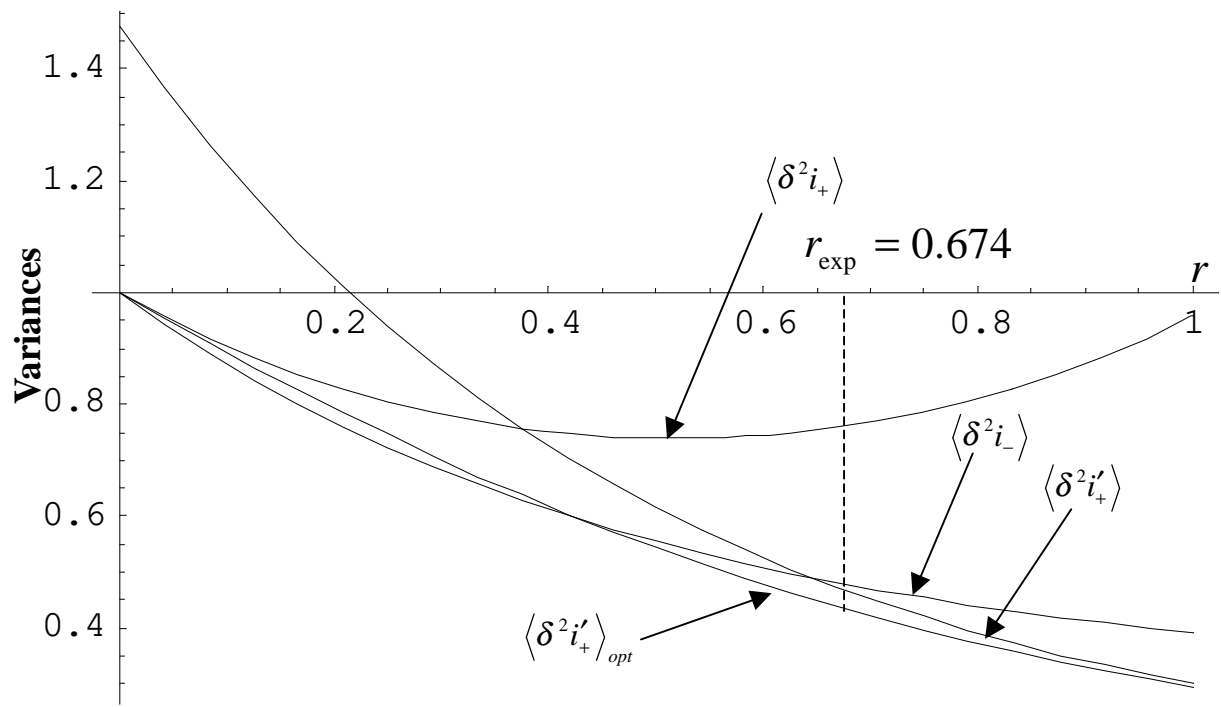
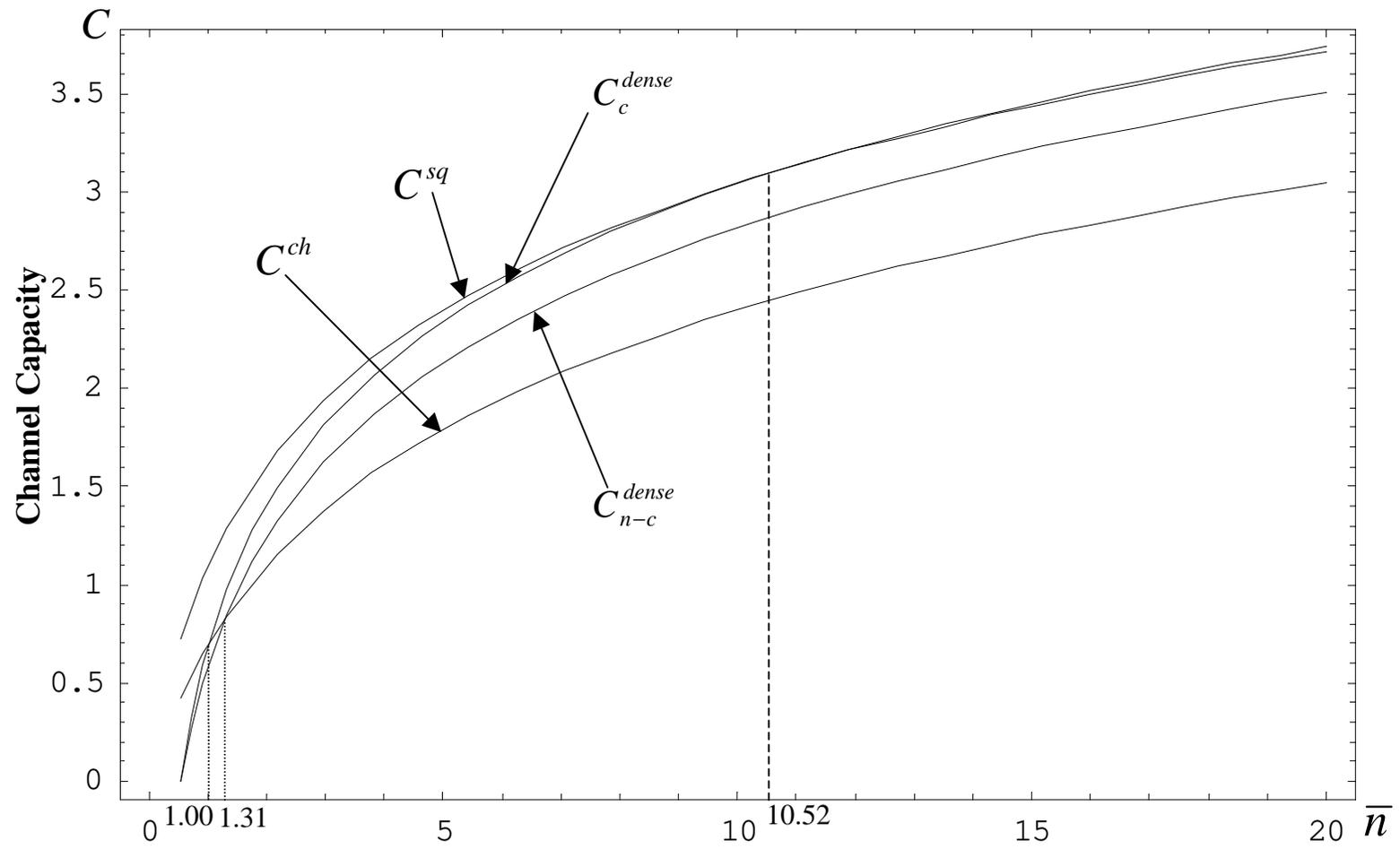
\end{document}